\documentclass[preprintnumbers, prd, showpacs, floatfix,onecolumn, superscriptaddress,nofootinbib]{revtex4}
\usepackage{amsfonts}
\usepackage{amsmath}
\usepackage{amssymb,epsf}
\usepackage{latexsym}
\usepackage{graphicx,epsfig}
\usepackage{amssymb}
\usepackage{subfigure}
\usepackage{epstopdf}

\begin{document}

\title{Critical behavior of Born-Infeld dilaton black holes}
\author{M. H. Dehghani $^{1,2} $\thanks{%
mhd@shirazu.ac.ir}, A. Sheykhi$^{1,2}$\thanks{%
asheykhi@shirazu.ac.ir} and Z. Dayyani}
\affiliation{Physics Department and Biruni Observatory, College of Sciences, Shiraz
University, Shiraz 71454, Iran\\
$^2$Research Institute for Astronomy and Astrophysics of Maragha (RIAAM),
P.O. Box 55134-441, Maragha, Iran}

\begin{abstract}
We explore the critical behavior of $(n+1)$-dimensional topological
Born-Infeld-dilaton black holes in an extended phase space. We treat the
cosmological constant and the Born-Infeld (BI) parameter as the
thermodynamic pressure and BI vacuum polarization which can vary. We obtain
thermodynamic quantities of the system such as pressure, temperature, Gibbs
free energy, and investigate the behaviour of these quantities. We also
study the analogy of the van der Waals liquid-gas system with the
Born-Infeld-dilaton black holes in canonical ensemble in which we can treat
the black hole charge as a fixed external parameter. Moreover, we show that
the critical values of pressure, temperature and volume are physical
provided the coupling constant of dilaton gravity is less than one and the
horizon is sphere. Finally, we calculate the critical exponents and show
that although thermodynamic quantities depend on the dilaton coupling
constant, BI parameter and the dimension of the spacetime, they are
universal and are independent of metric parameters.
\end{abstract}

\maketitle

\section{Introduction}

Recent works on the thermodynamics of black holes have shown that one may
enlarge the thermodynamic space to include the effective cosmological
constant and other parameters of the gravitational theory as thermodynamic
variables. For instance, in the case of Reissner-Nordstrom (RN) black hole
\cite{Do1}, considering the cosmological constant as a thermodynamic
variable proportional to the pressure: $P=-\Lambda /8\pi $, its conjugate
quantity will be the geometrical volume. In this case the black hole mass $M$
determines the enthalpy: $M=H\equiv U+PV$ which includes a contribution from
the energy of formation of the system \cite{Ka}. Also, from the point of
view of consistency of Smarr formula and first law of thermodynamics, one
should extend the thermodynamic space to include the cosmological constant
as a thermodynamic variable \cite{Do1}. The idea of including cosmological
constant as the thermodynamic pressure have been studied in many papers \cite%
{Ka,Do2,Do3,Ce1,Ur}. Although in the case of RN black holes the conjugate
quantity $V$ is the geometrical volume, it is not necessarily needed to be a
geometrical volume as it was revealed in the case of rotating black holes
\cite{Ce1}.

In addition to the extension of the thermodynamic space to include the
cosmological constant and its conjugate volume, one may extend this space to
include other parameters of a black hole provided the mass is equated to the
enthalpy. For instance, one may associate the non-geometrical thermodynamic
volumes to Taub--NUT, Taub--Bolt \cite{John} and Kerr-bolt \cite{Kerrbolt}
spacetimes. Also, considering nonlinear electrodynamics, one may extend the
thermodynamic space for the consistency of first law of thermodynamics with
the corresponding Smarr relation \cite{Mannb,Rabin,Zou,Hendi}. For instance
for Born-Infeld (BI) black holes, one should also consider the variation of
the minimal field strength $\beta $ in the first law to be consistent with
the corresponding Smarr relation \cite{Mannb}.

In this new context with extended thermodynamic space, one may study the
analogy between charged black holes in AdS space and the van der Waals fluid
and investigate the critical behaviour of the system. Phase transition and
critical behaviour in Einstein gravity have been investigated by many
authors \cite{Do1,Do2,Do3,Ur,Mannb,Rabin,Zou,Hendi,Mann1}. For the case of
Myers-Perry black holes, these have been investigated in \cite{Sherkat}.
Also, the critical behavior of higher order gravities such as Gauss-Bonnet
\cite{De,Sherkat1} and Lovelock gravity coupled to BI electrodynamics have
been investigated \cite{Xi}. The studies on the critical behavior of charged
black holes were also extended to dilaton gravity \cite{Ren}. In this
regards, the critical behavior of charged black holes of
Einstein-Maxwell-dilaton gravity in the presence of two Liouville-type
potentials which make the solution asymptotically neither flat nor AdS has
been explored in Ref. \cite{Kamrani}. It was found that the critical
exponents are universal and are independent of the details of the system
although the thermodynamic quantities depend on the dilaton coupling
constant \cite{Kamrani}.

In this paper we further generalize the studies on the extended
thermodynamic space and critical behavior of dilaton black holes by
investigating the critical behavior of the $(n+1)$-dimensional dilaton black
holes coupled to nonlinear BI electrodynamics \cite{BI} in an extended phase
space with fixed charge. Due to the fact that the BI Lagrangian coupled to a
dilaton field appears frequently in string theory, it is important to
investigate various properties of black hole solutions in this theory. As we
shall see the presence of the dilaton field affects the thermodynamic
properties of black holes. As in Ref. \cite{Mannb}, we consider the BI
parameter as a thermodynamic phase space variable to satisfy the Smarr
relation and introduce its conjugate quantity as polarization. We also
calculate the critical exponents and show that they are universal and are
independent of the nonlinearity parameter as well as the dilaton-
electromagnetic coupling constant.

This paper is organized as follows. In the next section we present the basic
field equations and consider a class of $(n+1)$-dimensional topological
black hole solutions in Einstein-Born-Infeld dilaton (EBId) gravity and
review their thermodynamic properties. In Sec. \ref{Ph} we study the phase
structure of the solution and present the generalized Smarr relation in the
presence of dilaton field. In Sec. \ref{state}, we obtain the equation of
state, study the critical behavior of the solutions and compare them with
van der Waals fluid. Gibbs free energy is investigated in Sec. \ref{Gibbs},
while critical exponents is considered in Sec. \ref{Exponent}. The last
section is devoted to summery and conclusion.

\section{Topological Born-Infeld-dilaton black holes in $(n+1)$ dimensions}

\label{BI}

The $(n+1)$-dimensional action in which gravity is coupled to a dilaton
field and nonlinear BI electrodynamics can be written as \cite{SheyBI}
\begin{equation}
S=\frac{1}{16}\int d^{n+1}x\sqrt{-g}\left( R-\frac{4}{n-1}\left( \nabla \Phi
\right) ^{2}-V\left( \Phi \right) +L\left( F,\Phi \right) \right) ,
\label{Action}
\end{equation}%
where $R$ is the Ricci scalar, $\Phi $ is the dilaton field, $V\left( \Phi
\right) $ is a potential for $\Phi $ and the BI Lagrangian $L\left( F,\Phi
\right) $ is given by
\begin{eqnarray}
L(F,\Phi ) &=&4\beta ^{2}e^{4\alpha \Phi /(n-1)}\mathcal{L}(Y) \\
\mathcal{L}(Y) &=&1-\sqrt{1+Y},  \label{LY} \\
Y &=&\frac{e^{-8\alpha \Phi /({n-1})}F^{2}}{2\beta ^{2}}.  \label{Y}
\end{eqnarray}%
In the above equations $F^{2}=F^{\mu \nu }F_{\mu \nu }$ with $F_{\mu \nu
}=\partial _{\mu }A_{\nu }-\partial _{\mu }A_{\nu }$ is the electromagnetic
field tensor with electromagnetic vector potential $A_{\mu }$ and $\alpha $
is the the coupling constant of the scalar and electromagnetic field. As the
BI parameter $\beta $ goes to infinity, $L\left( F,\Phi \right) $ reduces to
the standard Maxwell field coupled to dilaton.

The equations of motion can be obtained by varying the action (\ref{Action})
with respect to the gravitational field $g_{\mu \nu }$, the dilaton field $%
\Phi $ and the gauge field $A_{\mu }$ which yields the following field
equations
\begin{eqnarray}
\mathcal{R}_{\mu \nu }&=&\frac{4}{n-1}\left( \partial _{\mu }\Phi \partial
_{\nu }\Phi +\frac{1}{4}g_{\mu \nu }V(\Phi )\right) -4e^{-4\alpha \Phi/({n-1}%
)}\partial _{Y}\mathcal{L}(Y)F_{\mu \eta }F_{\nu }^{\text{ }\eta }  \notag \\
&&+\frac{4\beta^2}{n-1} e^{4\alpha \Phi/({n-1})}\left[ 2Y\partial _{Y}%
\mathcal{L}(Y)-\mathcal{L}(Y)\right] g_{\mu \nu },  \label{FE1}
\end{eqnarray}
\begin{equation}
\nabla ^{2}\Phi =\frac{n-1}{8}\frac{\partial V}{\partial \Phi }+2 \alpha
\beta ^{2} e^{4\alpha \Phi/({n-1})}\left[ 2{Y}\partial _{Y}\mathcal{L}(Y)-%
\mathcal{L}(Y)\right] ,  \label{FE2}
\end{equation}
\begin{equation}
\nabla _{\mu }\left(e^{-4\alpha \Phi/({n-1})}\partial _{Y}\mathcal{L}%
(Y)F^{\mu \nu }\right) =0.  \label{FE3}
\end{equation}
In particular, in the case of the linear electrodynamics with $\mathcal{L}%
(Y)=-{\frac{1}{2}}Y$, the system of equations (\ref{FE1})-(\ref{FE3}) reduce
to the well-known equations of Einstein-Maxwell dilaton gravity \cite%
{CHM,Shey}.

The most general $(n+1)$-dimensional static metric with constant curvature ($%
t=const.$, $r=const.$)-boundary may be written as
\begin{equation}
ds^{2}=-f(r)dt^{2}+\frac{dr^{2}}{f\left( r\right) }+r^{2}R^{2}\left(
r\right) d\Omega ^{2},  \label{Met1}
\end{equation}%
where $d\Omega ^{2}$ is an $\left( n-1\right) $-dimensional hypersurface
with constant curvature $\left( n-1\right) \left( n-2\right) k$ \ and volume
$\omega _{n-1}$. In general, one can set $k=0,1,-1$. The action (\ref{Action}%
) admits a static black hole solution with metric (\ref{Met1}) provided $%
V(\Phi )$ is taken as $V\left( \Phi \right) =2\Lambda _{0}e^{2\zeta _{0}\Phi
}+2\Lambda e^{2\zeta \Phi }$ and $R\left( r\right) =e^{2\alpha \Phi /\left(
n-1\right) }$, where $\Lambda _{0},\;\Lambda ,\;\zeta _{0}$ and $\zeta $ are
the following constants \cite{SheyBI}
\begin{equation}
\zeta _{0}=\frac{2}{\alpha \left( n-1\right) },\;\;\;\zeta =\frac{2\alpha }{%
n-1},\;\;\;\Lambda _{0}=-\frac{k\left( n-1\right) \left( n-2\right) \alpha
^{2}}{2b^{2}\left( 1-\alpha ^{2}\right) }.  \tag{ons}
\end{equation}%
With these assumptions the equations of motion (\ref{FE1})-(\ref{FE3}) admit
the following solution \cite{SheyBI}
\begin{eqnarray}
A_{t}(r) &=&\frac{qb^{(3-n)\gamma }}{\Upsilon r^{\Upsilon }} {_2F}_{1}\left( %
\left[ \frac{1}{2},\frac{\alpha ^{2}-2+n}{2n-2}\right] ,\left[ \frac{\alpha
^{2}+3n-4}{2n-2}\right] ,-\eta \right) , \\
\Phi \left( r\right) &=&\frac{\left( n-1\right) \alpha }{2\left( 1+\alpha
^{2}\right) }\ln \left( \frac{b}{r}\right) ,
\end{eqnarray}

\begin{eqnarray}
f\left( r\right) &=&-\frac{k\left( n-2\right) \left( \alpha ^{2}+1\right)
^{2}b^{-2\gamma }}{\left( \alpha ^{2}-1\right) \left( \alpha ^{2}+n-2\right)
}r^{2\gamma }-\frac{m}{r^{\left( n-1\right) \left( 1-\gamma \right) -1}}+%
\frac{2\Lambda \left( \alpha ^{2}+1\right) ^{2}b^{2\gamma }}{\left(
n-1\right) \left( \alpha ^{2}-n\right) }r^{2\left( 1-\gamma \right) }  \notag
\\
&&-\frac{4\beta ^{2}\left( \alpha ^{2}+1\right) ^{2}b^{2\gamma }r^{2\left(
1-\gamma \right) }}{\left( n-1\right) \left( \alpha ^{2}-n\right) }\left\{ 1-%
{_2F}_{1}\left( \left[ -\frac{1}{2},\frac{\alpha ^{2}-n}{2n-2}\right] ,\left[
\frac{n-2}{2n-2}\right] ,-\eta \right) \right\} ,
\end{eqnarray}%
where $b$ is an arbitrary nonzero positive constant, $\Lambda $ is a free
parameter which plays the role of the cosmological constant, $\gamma =\alpha
^{2}/(\alpha ^{2}+1)$ and
\begin{eqnarray*}
\Upsilon &=&(n-3)(1-\gamma )+1=\frac{n-2+\alpha ^{2}}{1+\alpha ^{2}} \\
\eta &=&\frac{q^{2}b^{2\gamma \left( 1-n\right) }}{\beta ^{2}r^{2\left(
n-1\right) \left( 1-\gamma \right) }}.
\end{eqnarray*}
In the above equations $m$ and $q$ are the mass and charge parameters,
respectively.

The ADM mass of the black hole is \cite{SheyBI}
\begin{equation}
M=\frac{b^{(n-1)\gamma }(n-1)\omega _{n-1}}{16\pi \left( \alpha
^{2}+1\right) }m,  \label{Mass}
\end{equation}%
where the mass parameter $m$ may be written in term of the horizon radius as
\begin{eqnarray}
m(r_{+}) &=&\frac{k(n-2)\left( 1+\alpha ^{2}\right) b^{-2\gamma }}{\left(
1-\alpha ^{2}\right) \Upsilon }r_{+}^{\Upsilon }-\frac{2\Lambda \left(
\alpha ^{2}+1\right) ^{2}b^{2\gamma }}{(n-1)\left( n-\alpha ^{2}\right) }%
r_{+}^{n(1-\gamma )-\gamma }  \notag \\
&&+\frac{4\beta ^{2}\left( \alpha ^{2}+1\right) ^{2}b^{2\gamma }}{\left(
n-1\right) \left( n-\alpha ^{2}\right) }r_{+}^{n(1-\gamma )-\gamma }\left\{
1-{_2F}_{1}\left( \left[ -\frac{1}{2},\frac{\alpha ^{2}-n}{2n-2}\right] ,%
\left[ \frac{\alpha ^{2}+n-2}{2n-2}\right] ,-\eta _{+}\right) \right\} .
\end{eqnarray}%
In the above equation, $r_{+}$ denotes the radius of the event horizon which
is the largest root of $f(r_{+})=0$ and $\eta _{+}$ is the value of $\eta $
at $r_{+}$. The temperature of the topological black hole on outer horizon $%
r_{+}$ can be written as%
\begin{equation}
T=\frac{f^{\prime }(r_{+})}{4\pi }=-\frac{\left( \alpha ^{2}+1\right)
b^{2\gamma }r_{+}^{1-2\gamma }}{2\pi \left( n-1\right) }\left( \frac{%
k(n-2)\left( \alpha ^{2}+1\right) ^{2}b^{-4\gamma }}{2\left( \alpha
^{2}-1\right) }r_{+}^{4\gamma -2}+\Lambda -2\beta ^{2}\left( 1-\sqrt{1+\eta
_{+}}\right) \right) .  \label{temp}
\end{equation}%
Using the so called area law of the entropy which states that the entropy of
the black hole is a quarter of the event horizon area, one can obtain
\begin{equation}
S=\frac{b^{(n-1)\gamma }r_{+}^{(n-1)(1-\gamma )}\omega _{n-1}}{4}.
\label{entropy}
\end{equation}%
Using Gauss law, the charge can be obtained as%
\begin{equation}
Q=\frac{q\omega _{n-1}}{4\pi },  \label{Q}
\end{equation}%
The electric potential $U$, measured at infinity with respect to the
horizon, is defined by
\begin{equation}
U=A_{\mu }\chi ^{\mu }\left\vert _{r\rightarrow \infty }-A_{\mu }\chi ^{\mu
}\right\vert _{r=r_{+}},
\end{equation}%
where $\chi =\partial _{t}$ is the null generator of the horizon. One
obtains \cite{SheyBI}
\begin{equation}
U=\frac{qb^{(3-n)\gamma }}{\Upsilon r_{+}{}^{\Upsilon }}{_2F}_{1}\left( %
\left[ \frac{1}{2},\frac{\alpha ^{2}-2+n}{2n-2}\right] ,\left[ \frac{\alpha
^{2}+3n-4}{2n-2}\right] ,-\eta \right) .  \label{U}
\end{equation}

\section{Phase Structure \label{Ph}}

In this section, we would like to investigate thermodynamics of BI-dilaton
black holes in an extended phase space in which the cosmological constant
and BI parameter and their conjugate quantities are treated as thermodynamic
variables. The conjugate quantity of the cosmological constant, which is
proportional to pressure, is volume. Using the fact that the entropy of
black hole is a quarter of the area of the horizon, the thermodynamic volume
$V$ is obtained as
\begin{equation}
V=\int 4Sdr_{+}=\frac{b^{(n-1)\gamma }r_{+}^{n-\gamma (n-1)}}{n-\gamma (n-1)}%
\omega _{n-1}.  \label{volume}
\end{equation}%
In the extended phase space $M$ should be a function of the extensive
quantities: entropy and charge, and intensive quantities: pressure, and
Born-Infeld parameter. Hence, defining $B$ as an extensive quantity
conjugate to $\beta $
\begin{equation}
B=\left( \frac{\partial M}{\partial \beta }\right) ,
\end{equation}%
the first law takes the form
\begin{equation}
dM=TdS+UdQ+VdP+Bd\beta .
\end{equation}%
It is easy to show that the conjugate quantities of the thermodynamic volume
and Born-Infeld parameter are
\begin{eqnarray}
P &=&-\frac{\Lambda }{8\pi }\frac{n-\gamma (n-1)}{n-\gamma (n+1)}\left(
\frac{b}{r_{+}}\right) ^{2\gamma }=-\frac{\left( n+\alpha ^{2}\right) }{%
\left( n-\alpha ^{2}\right) }\left( \frac{b}{r_{+}}\right) ^{2\gamma }\frac{%
\Lambda }{8\pi },  \label{press} \\
B &=&\frac{(1+\alpha ^{2})\beta b^{\gamma (n+1)}\omega _{n-1}}{2\pi
(n-\alpha ^{2})r^{\gamma (n+1)-n}}\left\{ 1-{_{2}F}_{1}\left( \left[ -\frac{1%
}{2},-\frac{n-\alpha ^{2}}{2(n-1)}\right] ,\left[ \frac{\alpha ^{2}+n-2}{%
2(n-1)}\right] ,-\eta _{+}\right) \right.  \notag \\
&&\left. -\frac{(n-\alpha ^{2})\eta _{+}}{2(n-2+\alpha ^{2})}{_{2}F}%
_{1}\left( \left[ \frac{1}{2},\frac{n+\alpha ^{2}-2}{2(n-1)}\right] ,\left[
\frac{\alpha ^{2}+3n-4}{2(n-1)}\right] ,-\eta _{+}\right) \right\} ,
\end{eqnarray}%
which shows that $B$ is an extensive quantity. One may note that pressure is
proportional to the cosmological constant $\Lambda $\textbf{,} while the
constant of proportionality depends on the dilaton parameter. We can see
that the above $P$ reduces to the pressure for Reissner-Nordstrom black hole
\cite{Mann1} or BI-AdS black hole \cite{Mannb} in the absence of dilaton ($%
\gamma =0=\alpha $). One may also note that the above expression for the
pressure is the same as that of Einstein-Maxwell dilaton black holes \cite%
{Kamrani}. Also, it is clear that the pressure is positive provided $\alpha <%
\sqrt{n}$. This is consistent with the argument given in \cite{SheyBI},
which state that the topological BI-dilaton black hole solutions exist
provided $\alpha <\sqrt{n}$. The Smarr relation may be obtained from the
values of thermodynamic variables and mass as

\begin{equation}
M=\frac{n-1}{n-2+\alpha ^{2}}TS-\frac{1-\alpha ^{2}}{n-2+\alpha ^{2}}\left(
2VP+\beta B\right) +UQ.
\end{equation}%
One may note that the above generalized Smarr formula reduces to those of
Refs. \cite{De,Sherkat1} in the absence of dilaton field ($\alpha =0$)%
\begin{equation}
M=\frac{n-1}{n-2}S\left( \frac{\partial M}{\partial S}\right) +Q\left( \frac{%
\partial M}{\partial Q}\right) -\frac{2}{n-2}P\left( \frac{\partial M}{%
\partial P}\right) -\frac{\beta }{n-2}\left( \frac{\partial M}{\partial
\beta }\right) .
\end{equation}%
In what follows, we study the phase transition of the charged BI-dilatonic
black hole system in the extended phase space in canonical ensemble. Indeed,
we treat the black hole charge $Q$ as a fixed external parameter, not a
thermodynamic variable.

\section{Equation of state \label{state}}

\begin{figure}[tbp]
\epsfxsize=8cm \centerline{\epsffile{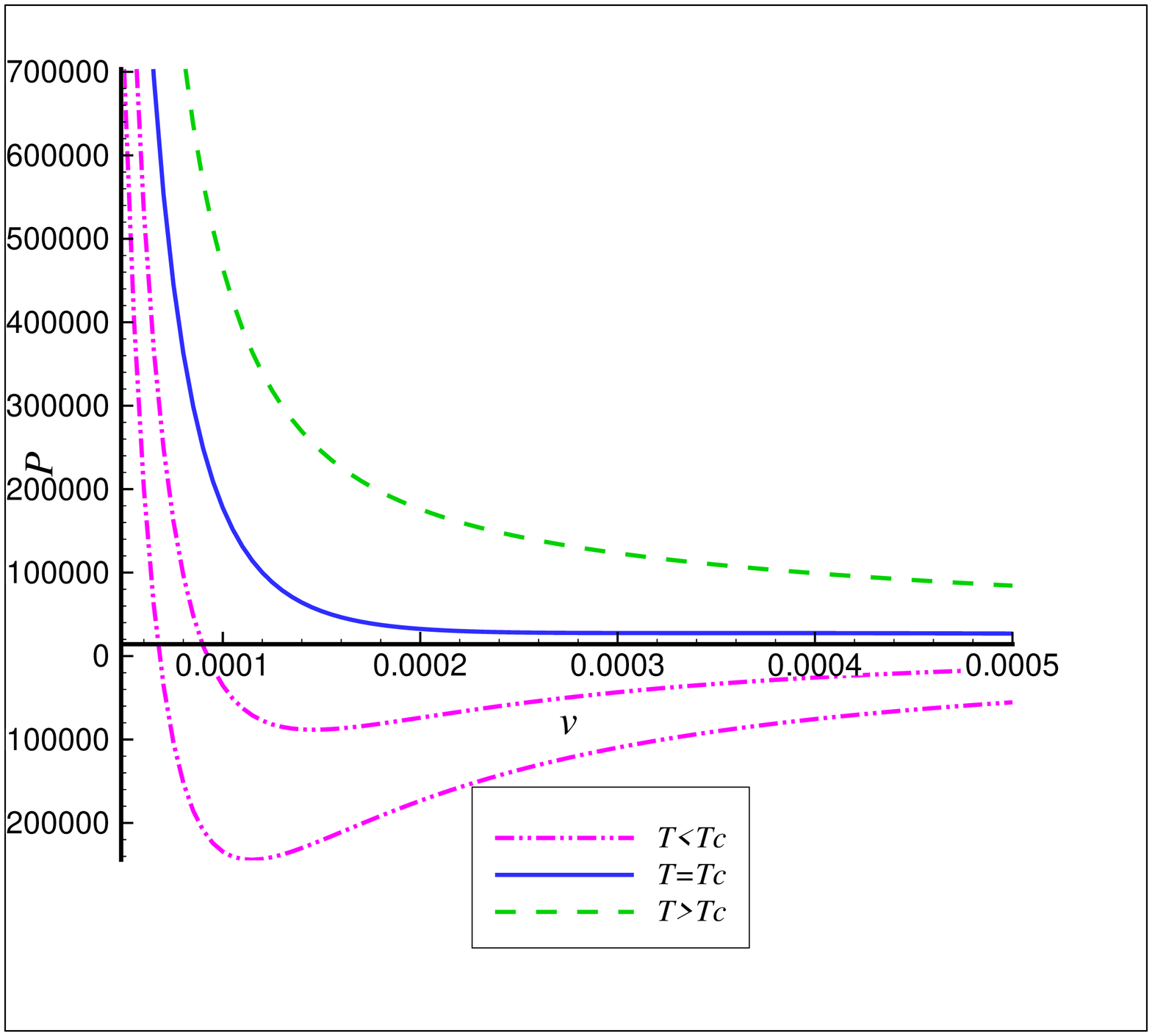}}
\caption{$P-v$ diagram of BID black holes for $b=1$, $n=3$, $q=1$, $k=1$, $%
\protect\beta =1$ and $\protect\alpha =0.3$.}
\label{Fig1}
\end{figure}

\begin{figure}[tbp]
\epsfxsize=8cm \centerline{\epsffile{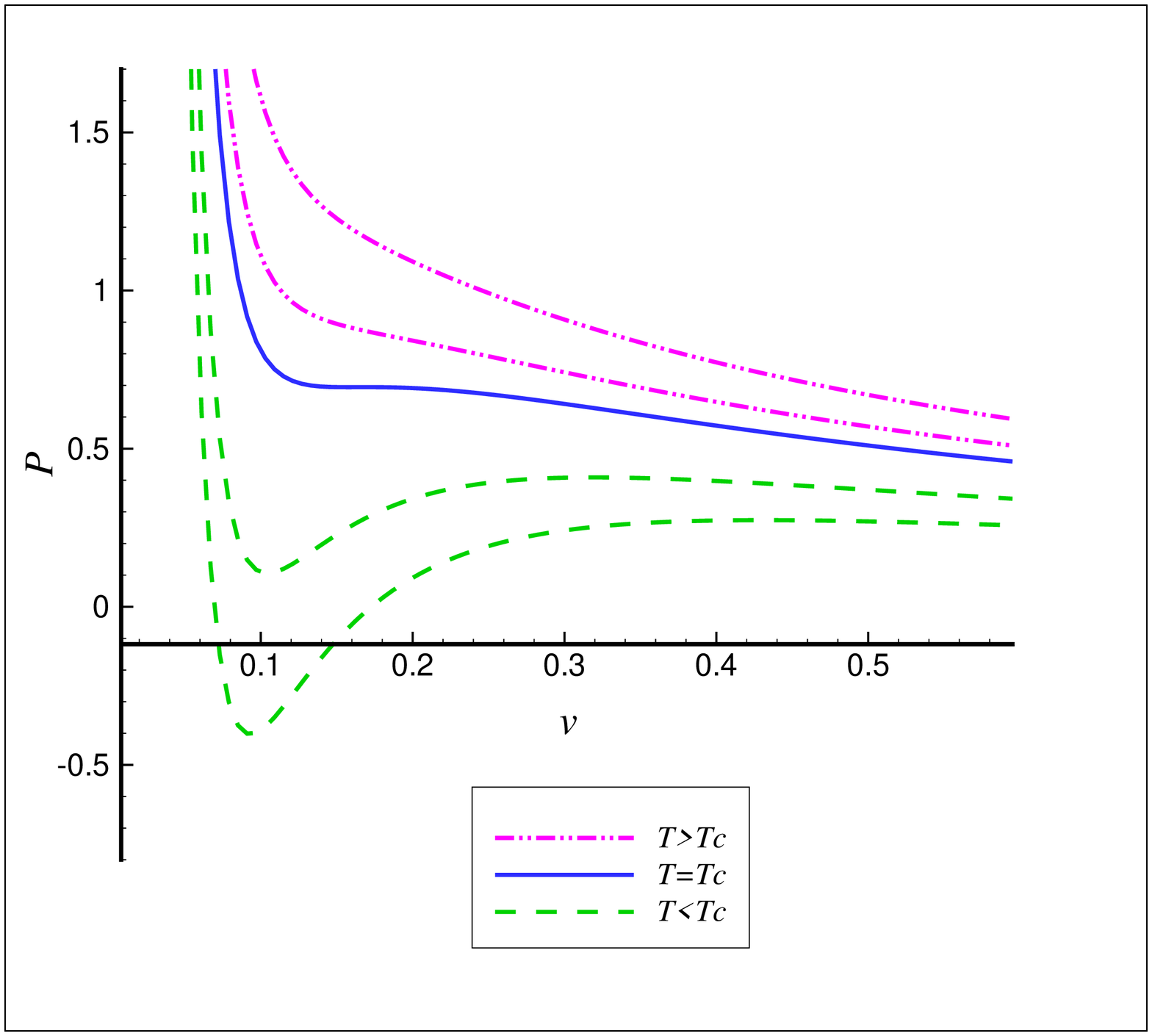}}
\caption{$P-v$ diagram of BID black holes for $b=1$, $n=3$, $q=1$, $k=1$, $%
\protect\beta =0.25$ and $\protect\alpha =0.3$.}
\label{Fig2}
\end{figure}

\begin{figure}[tbp]
\epsfxsize=8cm \centerline{\epsffile{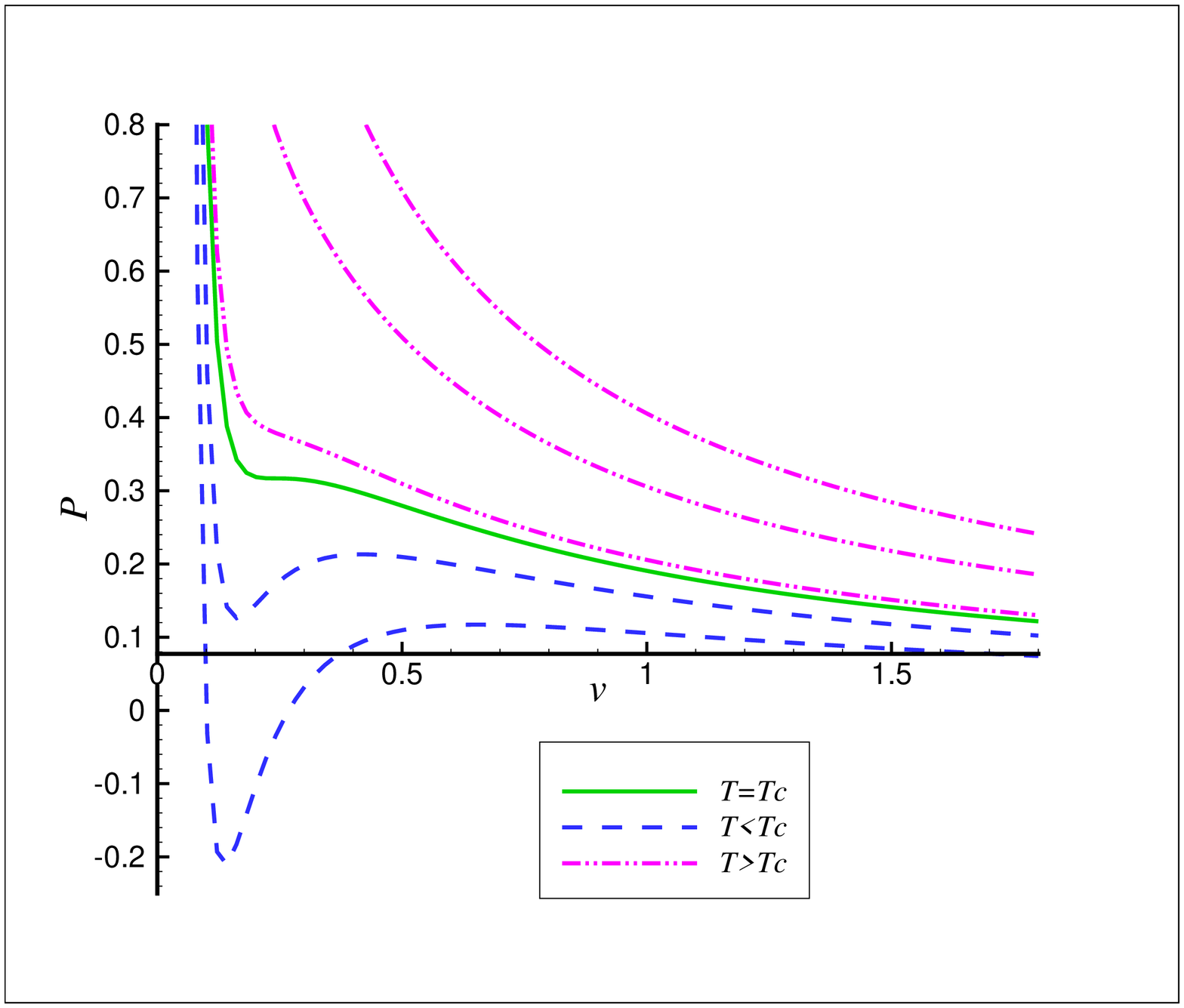}}
\caption{$P-v$ diagram of BID black holes for $b=1$, $n=3$, $q=1$, $k=1$, $%
\protect\beta =0.27$ and $\protect\alpha =0.3$.}
\label{Fig3}
\end{figure}

\begin{figure}[tbp]
\epsfxsize=8cm \centerline{\epsffile{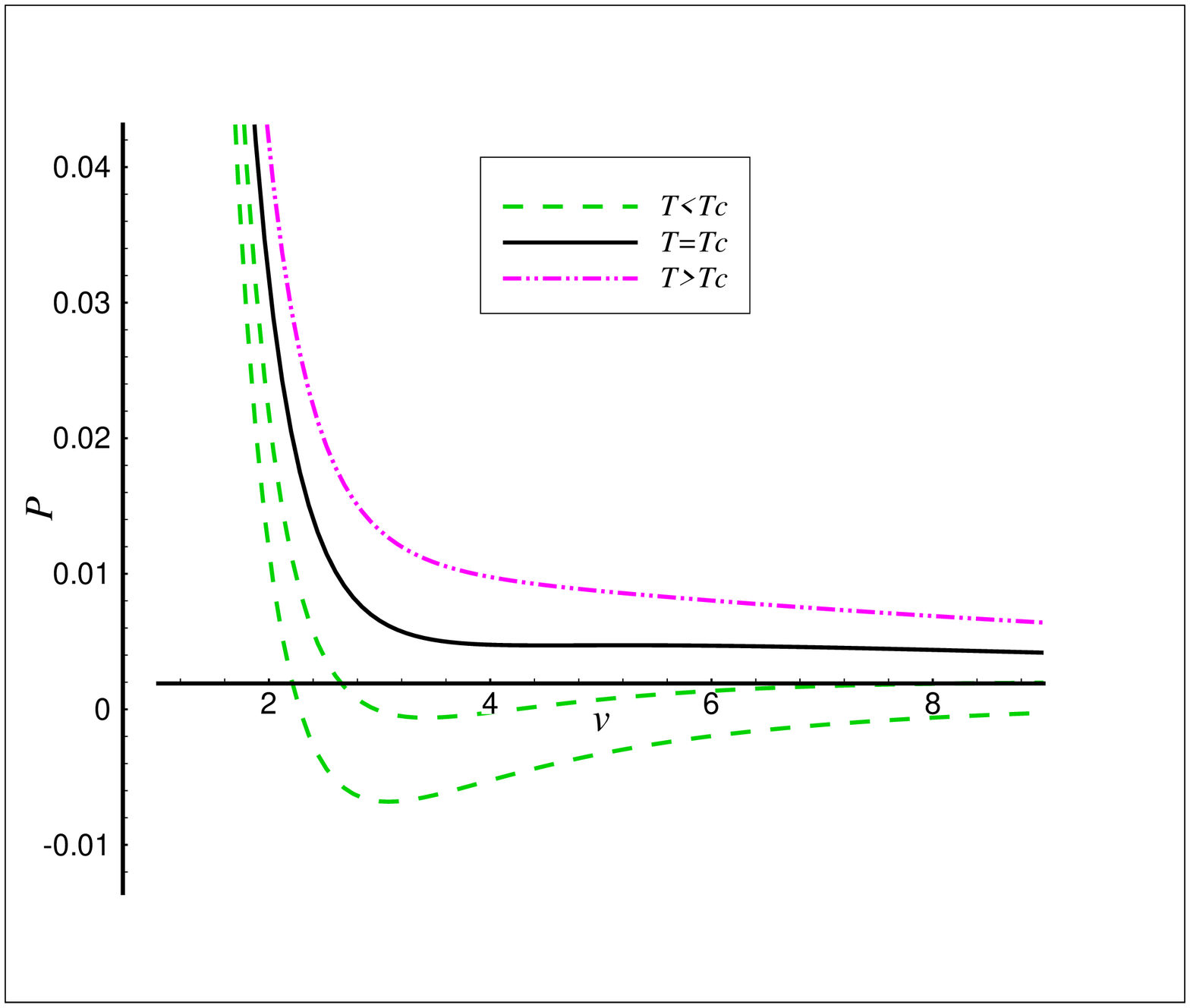}}
\caption{$P-v$ diagram of BID black holes for $b=1$, $n=3$, $q=1$, $k=1$, $%
\protect\beta =1$ and $\protect\alpha =0.3$.}
\label{Fig4}
\end{figure}

Using Eq. (\ref{press}) and regarding the charge $Q$ as a fixed parameter,
Eq. (\ref{temp}) can be written as
\begin{eqnarray}
P &=&\frac{\Gamma T}{r_{+}}-\frac{k(n-2)(1+\alpha ^{2})\Gamma }{4\pi
(1-\alpha ^{2})b^{2\gamma }r_{+}^{2-2\gamma }}  \notag \\
&&+\frac{\beta ^{2}(n+\alpha ^{2})b^{2\gamma }}{4\pi (n-\alpha
^{2})r_{+}^{2\gamma }}\left( \sqrt{1+\eta _{+}}-1\right) ,
\label{eq of state1}
\end{eqnarray}%
where
\begin{equation*}
\Gamma =\frac{\left( n-1\right) \left( n+\alpha ^{2}\right) }{4\left(
n-\alpha ^{2}\right) \left( \alpha ^{2}+1\right) }.
\end{equation*}%
Taking into account the fact that $r_{+}$ is a function of the thermodynamic
volume $V$, as one may see from Eq. (\ref{volume}), the above equation can
be regarded as the equation of state $P(V,T,\beta )$. Before proceeding
further, we translate the `geometric' equation of state (\ref{eq of state1})
to a physical one by performing a dimensional analysis. Noting that the
physical pressure and temperature are given by
\begin{equation}
\mathcal{P}=\frac{\hbar c}{l_{p}^{2}}P,\quad \mathcal{T}=\frac{\hbar c}{%
\kappa }T,
\end{equation}%
where the Planck length is $l_{p}=\sqrt{\hbar G/c^{3}}$ and $\kappa $ is the
Boltzmann constant, Eq. (\ref{eq of state1}) can be written as%
\begin{eqnarray}
\mathcal{P} &=&\frac{\kappa \Gamma \mathcal{T}}{l_{p}^{2}r_{+}}-\frac{%
k(n-2)(1+\alpha ^{2})\hbar c\Gamma }{4l_{p}^{2}\pi (1-\alpha ^{2})b^{2\gamma
}r_{+}^{2-2\gamma }}  \notag \\
&&+\frac{\beta ^{2}(n+\alpha ^{2})b^{2\gamma }\hbar c}{4\pi
l_{p}^{2}(n-\alpha ^{2})r_{+}^{2\gamma }}\left( \sqrt{1+\eta _{+}}-1\right) .
\label{Eqstateph}
\end{eqnarray}%
\newline
Now, comparing the above physical equation of state with the van der Walls
equation \cite{Mann1}%
\begin{equation*}
\mathcal{P}=\frac{\mathcal{T}}{v}+...,
\end{equation*}%
we understand that the specific volume $v$ of the fluid in terms of the
horizon radius should be written as,
\begin{equation}
v=\frac{l_{p}^{2}r_{+}}{\Gamma },  \label{volume2}
\end{equation}%
Returning to the geometrical units $(G=\hbar =c=1\Longrightarrow
l_{p}^{2}=1) $, the equation of state (\ref{eq of state1}) can be written as
\begin{eqnarray}
P &=&\frac{T}{v}-\frac{k(n-2)\left( \alpha ^{2}+1\right) \Gamma ^{^{2\gamma
-1}}}{4\pi \left( 1-\alpha ^{2}\right) b^{2\gamma }v^{2-2\gamma }}  \notag \\
&&+\frac{b^{2\gamma }\beta ^{2}\left( n+\alpha ^{2}\right) }{4\pi \left(
n-\alpha ^{2}\right) (v\Gamma )^{^{2\gamma }}}\left( \sqrt{1+\frac{(v\Gamma
)^{^{2\left( n-1\right) \left( \gamma -1\right) }}q^{2}}{b^{2\gamma \left(
n-1\right) }\beta ^{2}}}-1\right) .  \label{PvT}
\end{eqnarray}%
In order to compare the critical behavior of the system with van der Waals
gas, we should plot isotherm diagrams. The corresponding $P-v$ diagrams are
displayed in Figs. \ref{Fig1}-\ref{Fig4}.

The behavior of the isotherms diagrams depend on how deep we are in BI
nonlinear regime. The critical point can be obtained by solving the
following equations
\begin{equation}
\frac{\partial P}{\partial v}\Big|_{T_{c}}=0,\quad \frac{\partial ^{2}P}{%
\partial v^{2}}\Big|_{T_{c}}=0.  \label{CritEq}
\end{equation}

\subsection{Large $\protect\beta$}

First, we consider the the case of large $\beta $. In this case, Eq. (\ref%
{CritEq}) leads to
\begin{eqnarray}
v_{c} &=&\left( \frac{X}{b^{2(3-n)\gamma }}\right) ^{-1/(2\Upsilon )}\frac{%
q^{1/\Upsilon }}{\Gamma }\left\{ 1-\frac{(\alpha ^{2}+1)(\alpha
^{2}+4n-5)(\alpha ^{2}+2n-2)}{4k(n+\alpha ^{2}-2)(n-1)(n-2)\beta ^{2}}%
q^{(4\gamma -2)/\Upsilon }\frac{X^{[2(\Upsilon -\gamma )+1]/\Upsilon }}{%
b^{2(n-1)\gamma /\Upsilon }}+O(\frac{1}{\beta ^{4}})\right\} , \\
P_{c} &=&\frac{\Gamma X^{(\Upsilon -\gamma +1)/\Upsilon }}{2\pi (n-1)}%
\left\{ \frac{(2n-3+\alpha ^{2})(n-2+\alpha ^{2})}{b^{2\gamma /\Upsilon
}q^{2(1-\gamma )/\Upsilon }}+\frac{(\alpha ^{2}+4n-5)X^{(\Upsilon -2\gamma
+1)/\Upsilon }}{4\beta ^{2}b^{2n\gamma /\Upsilon }q^{2(2-3\gamma )/\Upsilon }%
}+O(\frac{1}{\beta ^{4}})\right\} , \\
T_{c} &=&\frac{(\alpha ^{2}+n-2)k(n-2)X^{(1-2\gamma )/(2\Upsilon )}}{\pi
(1-\alpha ^{2})(2n-3+\alpha ^{2})b^{\gamma (n-1)/\Upsilon }q^{(1-2\gamma
)/\Upsilon }}+\frac{(\alpha ^{2}+2n-2)X^{[2(2n-3)(1-\gamma )+1]/2\Upsilon }}{%
4\pi (n-1)b^{3\gamma (n-1)/\Upsilon }\beta ^{2}q^{3(1-2\gamma )/\Upsilon }}%
+O(\frac{1}{\beta ^{4}}),
\end{eqnarray}%
where
\begin{equation}
X=\frac{k(n-1)(n-2)}{2(2n-3+\alpha ^{2})\left( \alpha ^{2}+n-1\right) }.
\end{equation}%
It is easy to show the above critical quantities reduces to \cite{Kamrani}
as $\beta \rightarrow \infty $. Using the above critical values, $\rho _{c}$
is obtained as
\begin{equation}
\rho _{c}=\frac{P_{c}v_{c}}{T_{c}}=\frac{(1-\alpha ^{2})(2n-3+\alpha ^{2})}{%
4(n-1+\alpha ^{2})}\left\{ 1-\frac{\left( n-1\right) \left( 1+\alpha
^{2}\right) \left( b^{-2\alpha ^{2}}X\right) ^{(n-1)(1-\gamma )/\Upsilon }}{%
4\beta ^{2}\left( \alpha ^{2}+n-2\right) \left( \alpha ^{2}+n-1\right)
q^{2(1-2\gamma )/\Upsilon }}+O(\frac{1}{\beta ^{4}})\right\} ,
\label{universal ratio}
\end{equation}%
As one expects, the above $\rho _{c}$ reduces to that of Ref. \cite{Kamrani}
as $\beta \rightarrow \infty $,
\begin{equation}
\rho _{c}=\frac{P_{c}v_{c}}{T_{c}}=\frac{(1-\alpha ^{2})(2n-3+\alpha ^{2})}{%
4(n-1+\alpha ^{2})},
\end{equation}%
and in the absence of the dilaton field ($\alpha =0=\gamma $) in four
dimensions ($n=3$), it reduces to $3/8$ which is the characteristic of van
der Waals fluid. One should note that $\rho _{c}$ is positive and real
provided $\alpha <1$.

\subsection{Small $\protect\beta $}

In deep BI nonlinear regime, Eq. (\ref{PvT}) can be written as%
\begin{equation}
P=\frac{T}{v}-\frac{k(n-2)\left( \alpha ^{2}+1\right) \Gamma ^{^{2\gamma -1}}%
}{4\pi \left( 1-\alpha ^{2}\right) b^{2\gamma }v^{2-2\gamma }}+\frac{\left(
n+\alpha ^{2}\right) \beta q(\Gamma v)^{-n+1+(n-3)\gamma }}{4\pi \left(
n-\alpha ^{2}\right) b^{(n-3)\gamma }}.  \label{Psmall}
\end{equation}
Using Eq. (\ref{CritEq}) one can obtain the critical values as%
\begin{eqnarray}
v_{c} &=&\frac{b^{\alpha ^{2}(5-n)/(n-3+2\alpha ^{2})}}{\Gamma Z^{(1+\alpha
^{2})/(n-3+2\alpha ^{2})}}\beta ^{(1+\alpha ^{2})/(n-3+2\alpha ^{2})}, \\
P_{c} &=&\frac{(2\alpha ^{2}+n-3)(n+\alpha ^{2})(\alpha
^{2}+n-2)qZ^{(n-1+2\alpha ^{2})/(n-3+2\alpha ^{2})}}{8\pi (n-\alpha
^{2})(1+\alpha ^{2})b^{4\alpha ^{2}/(n-3+2\alpha ^{2})}\beta
^{2/(n-3+2\alpha ^{2})}}, \\
T_{c} &=&\frac{{(2\alpha ^{2}+n-1)}{(2\alpha ^{2}+n-3)}qZ^{(n-2+\alpha
^{2})/(n-3+2\alpha ^{2})}}{\pi (n-1)(1-\alpha ^{2})b^{(n-1)\alpha
^{2}/(n-3+2\alpha ^{2})}\beta ^{(1-\alpha ^{2})/(n-3+2\alpha ^{2})}},
\end{eqnarray}%
where
\begin{equation}
Z=\frac{k(n-1)(n-2)}{2q(\alpha ^{2}+n-2)(2\alpha ^{2}+n-1)}
\end{equation}%
which leads to the following value for $\rho _{c}$%
\begin{equation}
\rho _{c}=\frac{(n-2+\alpha ^{2})(1-\alpha ^{2})}{2(n-1+2\alpha ^{2})}.
\end{equation}%
One should note that $\rho _{c}$ in deep Born-Infeld regime is
different from that in the presence of Maxwell field. Also as in the case of
large $\beta $, the first order approximation of $\rho _{c}$ does not depend
on $\beta $, while upon inclusion of higher power of $\beta $ we expect this
to become $\beta $ dependent. Also, one should note that as $\beta $ goes to
zero $T_{c}$ or $P_{c}$ goes to infinity. Indeed, at $\beta =0$ the third
term in Eq. (\ref{Psmall}) vanishes and no critical behavior will occur.

\section{Gibbs free energy \label{Gibbs}}

\begin{figure}[tbp]
\epsfxsize=8cm \centerline{\epsffile{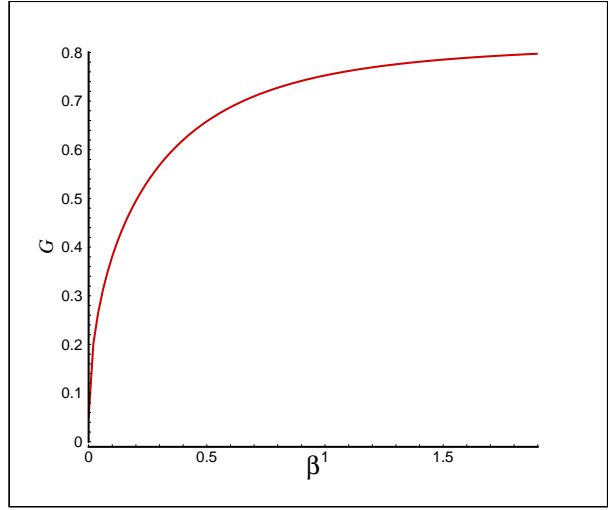}}
\caption{Gibbs free energy versus $\protect\beta $ for $b=1$, $n=3$, $q=1$, $%
k=1$, p=0.1, r=1, and $\protect\alpha =0.3$.}
\label{Fig10}
\end{figure}

In the canonical ensemble with fixed charge, the potential, which is the
free energy of the system presents the thermodynamic behaviour of a system
in a standard approach. But, since we are considering an extended phase
space, we associate it with the Gibbs free energy $G=M-TS$ \cite{Do1}. The
Gibbs free energy can be obtained as
\begin{eqnarray}
G &=&G\left( T,P\right) =\frac{(1+\alpha ^{2})b^{(n-1)\gamma }\omega _{n-1}}{%
4\pi r_{+}^{n(\gamma -1)-\gamma }}\left\{ \frac{k(n-2)b^{-2\gamma }}{%
4(n-2+\alpha ^{2})r_{+}^{2(1-\gamma )}}-\frac{4\pi P(1-\alpha ^{2})}{%
(n-1)(n+\alpha ^{2})}\right.   \notag \\
&&\left. -\frac{\beta ^{2}b^{2\gamma }}{(n-\alpha ^{2})r^{2\gamma }}\left\{ {%
_{2}F}_{1}\left( \left[ -\frac{1}{2},-\frac{n-\alpha ^{2}}{2n-2}\right] ,%
\left[ \frac{\alpha ^{2}+n-2}{2n-2}\right] ,-\eta _{+}\right) -1\right\}
\right.   \notag \\
&&\left. +\frac{b^{2\gamma }\beta ^{2}}{(n-1)r^{2\gamma }}\left( \sqrt{%
1+\eta _{+}}-1\right) \right\} ,  \label{Gibb}
\end{eqnarray}%
where $r_{+}$ is understood as a function of pressure and temperature via
equation of state (\ref{eq of state1}). If we expand the Gibbs free energy
for large $\beta $, we arrive at
\begin{eqnarray}
G\left( T,P\right)  &=&\frac{b^{(n-1)\gamma }\omega _{n-1}}{4\pi
r_{+}^{n(\gamma -1)-\gamma }}\left\{ \frac{k(n-2)(n-2+\alpha ^{2})}{%
4(1+\alpha ^{2})b^{2\gamma }r_{+}^{2(1-\gamma )}}-\frac{4\pi P(1-\alpha ^{4})%
}{(n-1)(n+\alpha ^{2})}\right.   \notag \\
&&\left. +\frac{q^{2}(2n-3+\alpha ^{2})(\alpha ^{2}+1)b^{-2(n-2)\gamma }}{%
2(n-2+\alpha ^{2})(n-1)r_{+}^{2(\Upsilon -\gamma +1)}}+O\left( \frac{1}{%
\beta ^{2}}\right) \right\} ,
\end{eqnarray}%
which reduces to the result obtained for black holes in EMd gravity as $%
\beta $ goes to infinity \cite{Kamrani}. Although the
hypergeometrical series expression in Eq. (\ref{Gibb})  is convergent only
for $\eta _{+}<1$, one may use the integral representation of
hypergeometrical function for any value $\eta _{+}$. In this case
one can obtain the limit of Gibbs free energy as $\beta $\ goes to
zero as
\begin{equation*}
\lim_{\beta \rightharpoonup 0}G=\frac{(1+\alpha ^{2})b^{(n-1)\gamma }\omega
_{n-1}}{4\pi r_{+}^{n(\gamma -1)-\gamma }}\left\{ \frac{k(n-2)b^{-2\gamma }}{%
4(n-2+\alpha ^{2})r_{+}^{2(1-\gamma )}}-\frac{4\pi P(1-\alpha ^{2})}{%
(n-1)(n+\alpha ^{2})}\right\} .
\end{equation*}%
Thus, the Gibbs free energy starts from the above value and increases to
\begin{eqnarray}
\lim_{\beta \rightharpoonup \infty }G &=&\frac{b^{(n-1)\gamma }\omega _{n-1}%
}{4\pi r_{+}^{n(\gamma -1)-\gamma }}\left\{ \frac{k(n-2)(n-2+\alpha ^{2})}{%
4(1+\alpha ^{2})b^{2\gamma }r_{+}^{2(1-\gamma )}}-\frac{4\pi P(1-\alpha ^{4})%
}{(n-1)(n+\alpha ^{2})}\right.   \notag \\
&&\left. +\frac{q^{2}(2n-3+\alpha ^{2})(\alpha ^{2}+1)b^{-2(n-2)\gamma }}{%
2(n-2+\alpha ^{2})(n-1)r_{+}^{2(\Upsilon -\gamma +1)}}\right\}
\end{eqnarray}%
as $\beta $ goes from zero to infinity. This can be seen in Fig. \ref{Fig10}%
.

\begin{figure}[tbp]
\epsfxsize=8cm \centerline{\epsffile{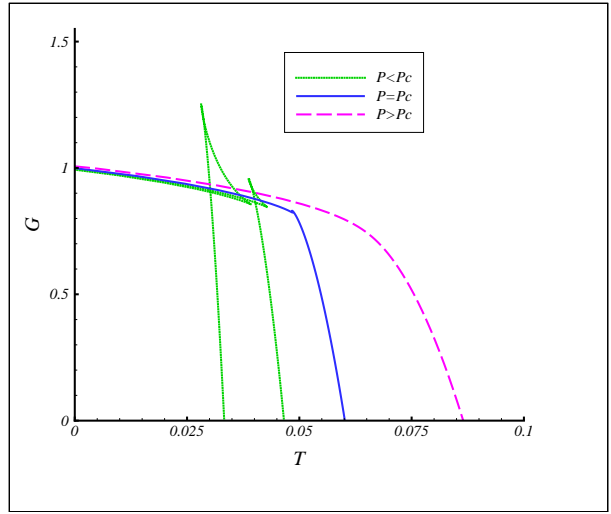}}
\caption{Gibbs free energy versus $T$ for $b=1$, $n=3$, $q=1$, $k=1$, $%
\protect\beta =1$ and $\protect\alpha =0.2$.}
\label{Fig1F}
\end{figure}

\begin{figure}[tbp]
\epsfxsize=8cm \centerline{\epsffile{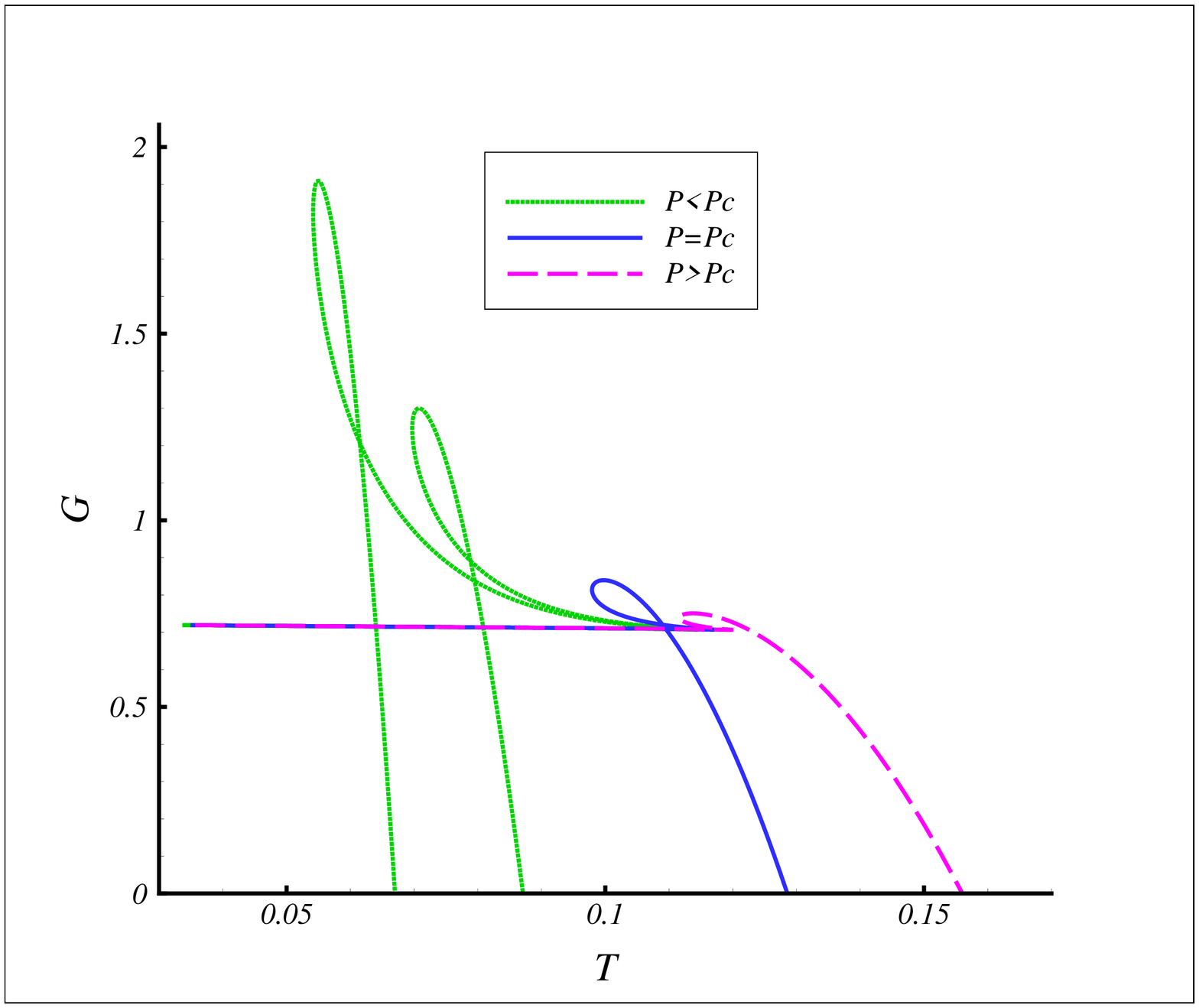}}
\caption{Gibbs free energy versus $T$ for $b=1$, $n=3$, $q=1$, $k=1$, $%
\protect\beta =0.25$ and $\protect\alpha =0.5$.}
\label{Fig2F}
\end{figure}

\begin{figure}[tbp]
\epsfxsize=8cm \centerline{\epsffile{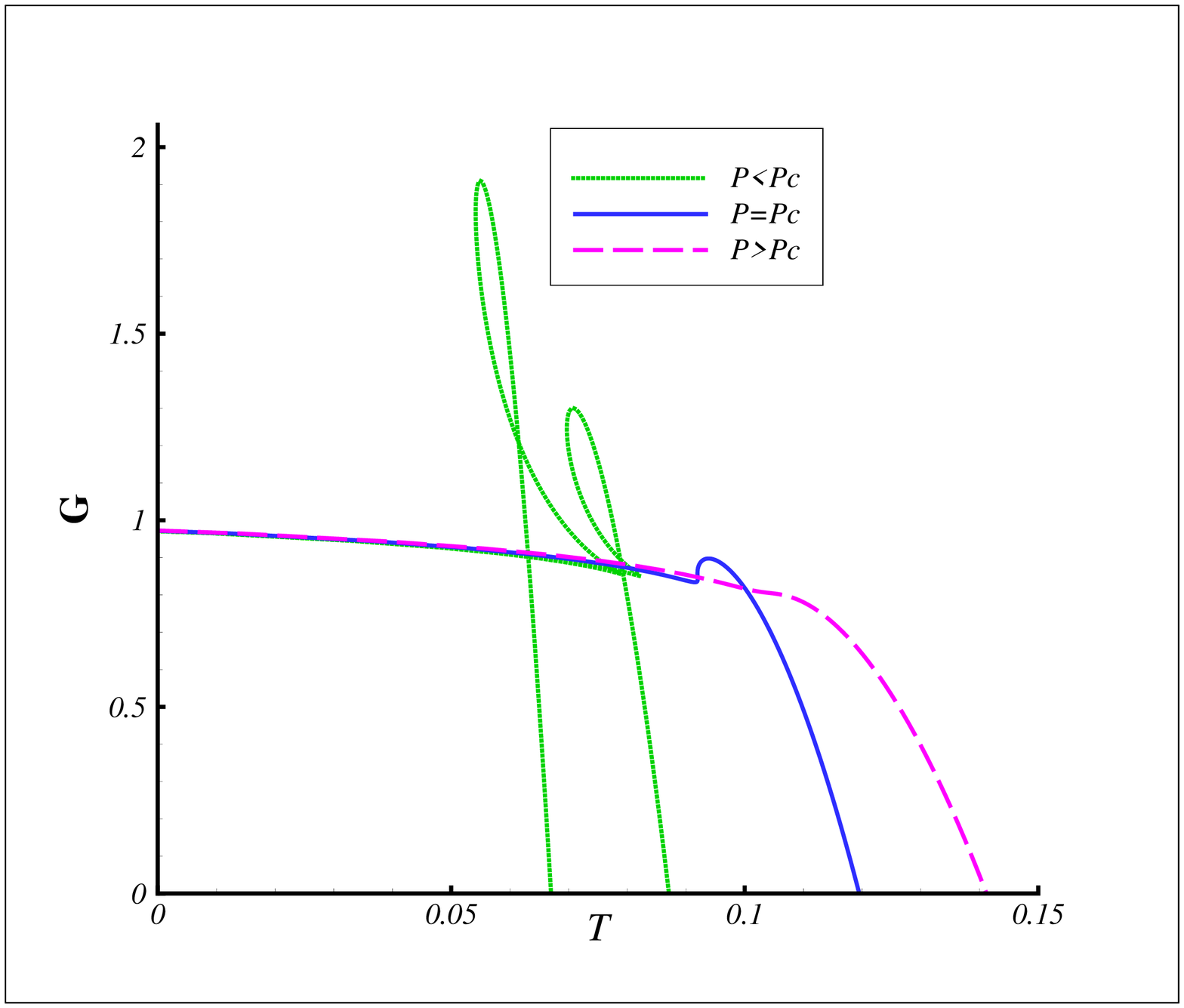}}
\caption{Gibbs free energy versus $T$ for $b=1$, $n=3$, $q=1$, $k=1$, $%
\protect\beta =0.5$ and $\protect\alpha =0.5$.}
\label{Fig3F}
\end{figure}

\begin{figure}[tbp]
\epsfxsize=8cm \centerline{\epsffile{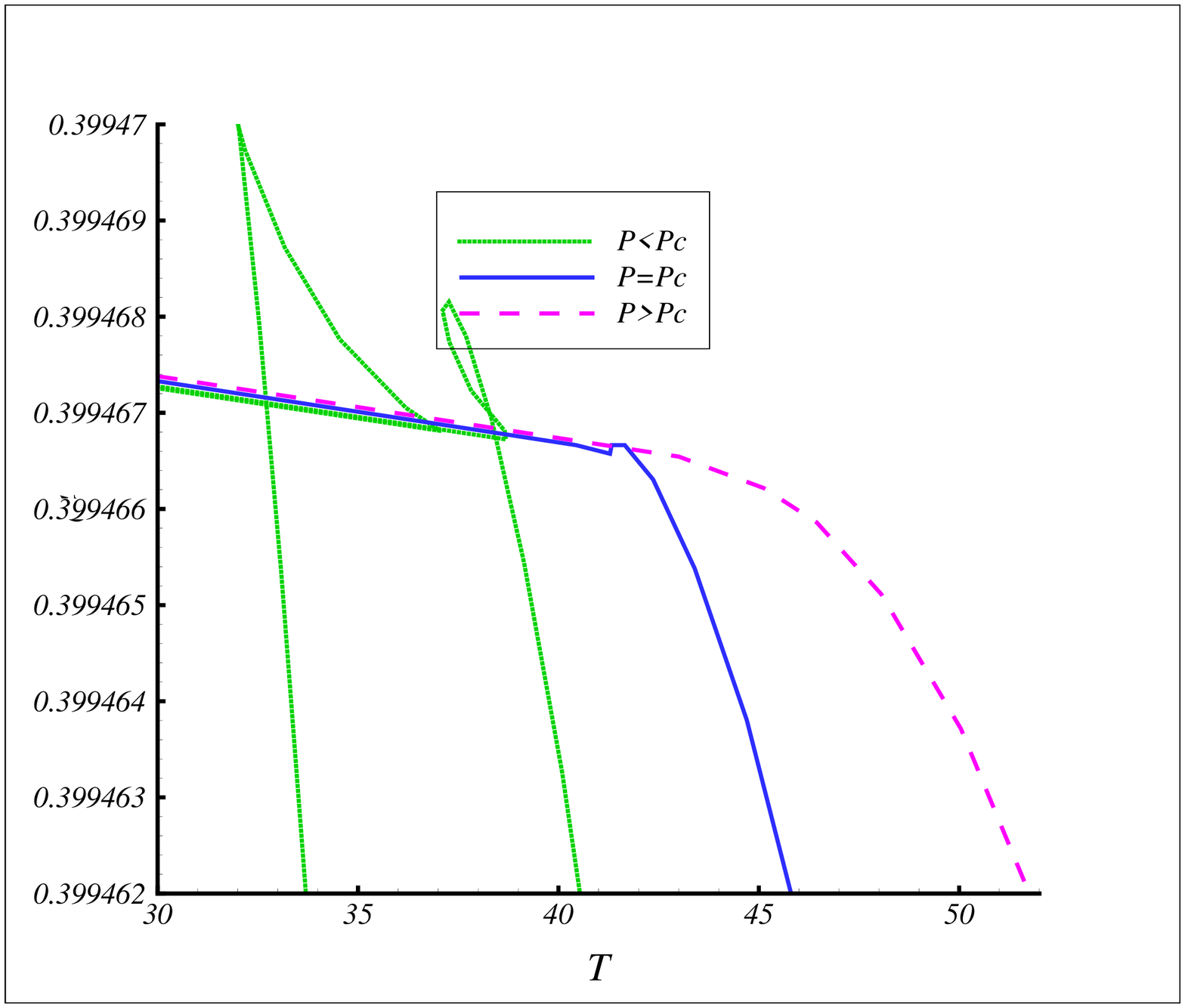}}
\caption{Gibbs free energy versus $T$ for $b=1$, $n=3$, $q=1$, $k=1$, $%
\protect\beta =0.1$ and $\protect\alpha =0.2$.}
\label{Fig4F}
\end{figure}

\begin{figure}[tbp]
\epsfxsize=8cm \centerline{\epsffile{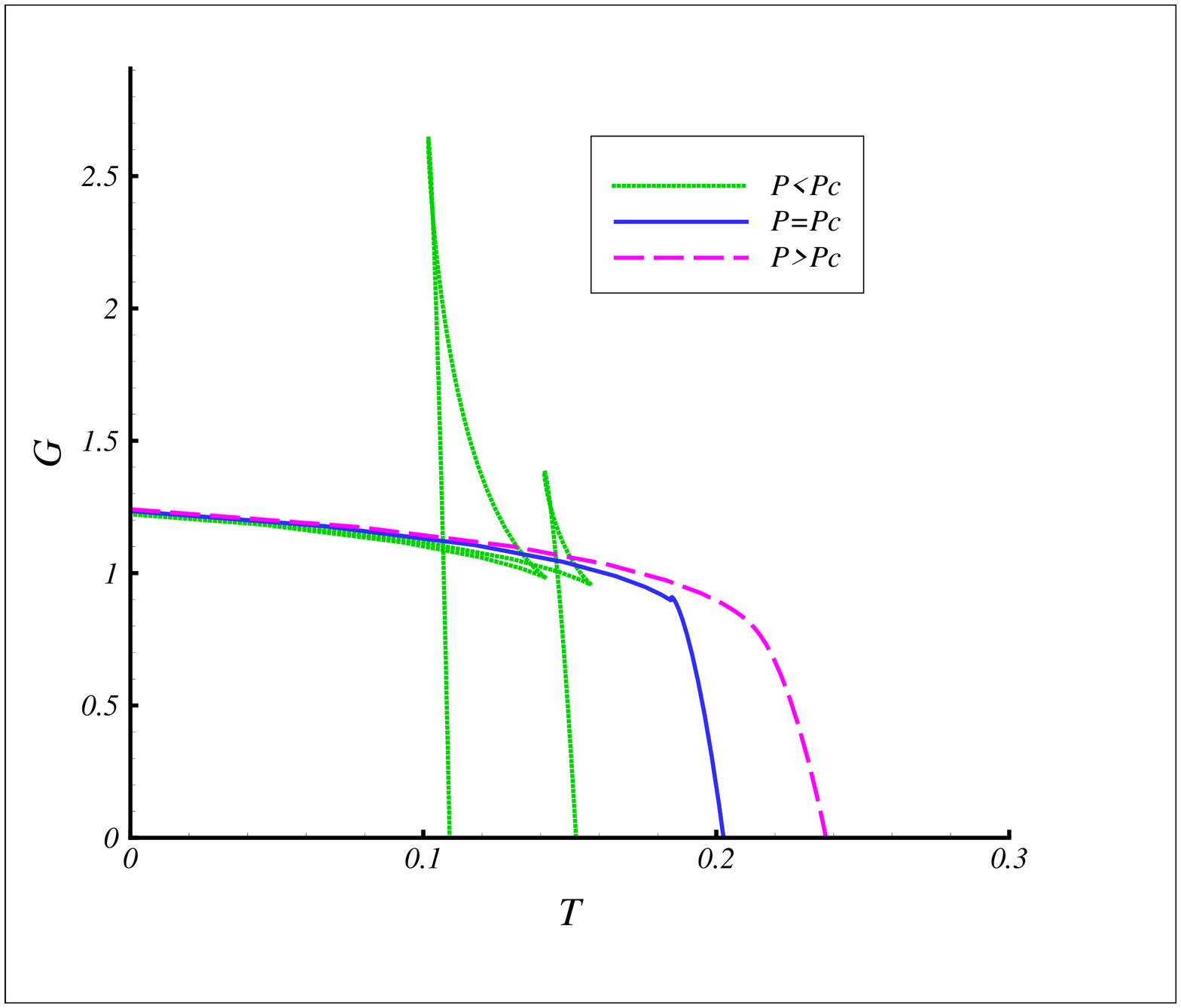}}
\caption{Gibbs free energy versus $T$ for $b=1$, $n=4$, $q=1$, $k=1$, $%
\protect\beta =1$ and $\protect\alpha =0.2$.}
\label{Fig5F}
\end{figure}

The behavior of the Gibbs free energy in terms of temperature is shown in
Figs. \ref{Fig1F}-\ref{Fig5F}. From these figures we see that there is a
swallowtail behavior for Gibbs free energy as a function of temperature,
which means that we have a first order small-large black hole transition for
the system.

\section{Critical exponents \label{Exponent}}

The behavior of physical quantities in the vicinity of critical point can be
characterized by the critical exponents. So, following the approach of \cite%
{Mannb}, one can calculate the critical exponents $\alpha ^{\prime }$, $%
\beta ^{\prime }$, $\gamma ^{\prime }$ and $\delta ^{\prime }$ for the phase
transition of an $(n+1)$-dimensional charged dilatonic black hole in the
presence of BI field. To obtain the critical exponents, we define the
reduced thermodynamic variables as
\begin{equation*}
p=\frac{P}{P_{c}},\quad \nu =\frac{v}{v_{c}},\quad \tau =\frac{T}{T_{c}}.
\end{equation*}%
So, equation of state (\ref{PvT})\textbf{\ }translates into the law of
corresponding state,
\begin{eqnarray}
p &=&\frac{1}{\rho _{c}}\frac{\tau }{\nu }-\frac{k(n-2)\left( \alpha
^{2}+1\right) \Gamma ^{^{2\gamma -1}}v_{c}^{2\gamma -2}}{4\pi P_{c}\left(
1-\alpha ^{2}\right) b^{2\gamma }\nu ^{2-2\gamma }}  \notag \\
&&+\frac{(\nu \Gamma v_{c})^{^{-2\gamma }}b^{2\gamma }\beta ^{2}\left(
n+\alpha ^{2}\right) }{4\pi P_{c}\left( n-\alpha ^{2}\right) }\left( \sqrt{1+%
\frac{(\nu \Gamma v_{c})^{^{2\left( n-1\right) \left( \gamma -1\right)
}}q^{2}}{b^{2\gamma \left( n-1\right) }\beta ^{2}}}-1\right) .  \label{law}
\end{eqnarray}%
which reduces to
\begin{equation*}
p=\frac{1}{\rho _{c}}\frac{\tau }{\nu }-\frac{k(n-2)\left( \alpha
^{2}+1\right) \Gamma ^{^{2\gamma -1}}}{4\pi b^{2\gamma }P_{c}\left( 1-\alpha
^{2}\right) (\nu v_{c})^{2-2\gamma }}+\frac{\left( n+\alpha ^{2}\right)
q^{2}(\nu \Gamma v_{c})^{^{2\left( n-1\right) \left( \gamma -1\right) }}}{%
8\pi P_{c}\left( n-\alpha ^{2}\right) b^{2\gamma (n-2)}(\nu \Gamma
)^{^{2\gamma }}}
\end{equation*}%
and
\begin{equation*}
p=\frac{1}{\rho _{c}}\frac{\tau }{\nu }-\frac{k(n-2)\left( \alpha
^{2}+1\right) \Gamma ^{^{2\gamma -1}}}{4\pi b^{2\gamma }P_{c}\left( 1-\alpha
^{2}\right) (\nu v_{c})^{2-2\gamma }}+\frac{\left( n+\alpha ^{2}\right)
\beta q(\nu \Gamma v_{c})^{-n+1+(n-3)\gamma }}{4\pi P_{c}\left( n-\alpha
^{2}\right) b^{(n-3)\gamma }}
\end{equation*}%
for large and small $\beta $, respectively. Although this law depends on
parameter $\gamma $, this doesn't affect the behavior of the critical
exponents as we will see bellow. To calculate the critical exponent $\alpha
^{\prime }$, we consider the entropy $S$ (\ref{entropy}) as a function of $T$
and $V$. Using Eq. (\ref{volume}) we have
\begin{equation*}
S=S\left( T,V\right) =\frac{b^{(n-1)\gamma }\omega _{n-1}}{4}\left\{ \frac{%
\left[ (n-1)(1-\gamma )+1\right] V}{b^{(n-1)\gamma }\omega _{n-1}}\right\}
^{(n-1)/(n+\alpha ^{2})}.
\end{equation*}%
Obviously, this is independent of $T$ and therefore the specific heat
vanishes, $C_{V}=T\left( \partial S/\partial T\right) _{V}=0$. Since the
exponent $\alpha ^{\prime }$ governs the behavior of the specific heat at
constant volume $C_{V}\varpropto \left\vert \tau -1\right\vert ^{\alpha
^{\prime }}$, hence the exponent\ $\alpha ^{\prime }=0$. Expanding Eq. (\ref%
{law}) near the critical point
\begin{equation}
\tau =t+1,\quad \nu =\left( \omega +1\right) ^{\frac{1}{\varepsilon }},
\end{equation}%
where\ $\varepsilon $ is a positive parameter defined as $\varepsilon
=n-\gamma \left( n-1\right) =(n+\alpha ^{2})/(1+\alpha ^{2})$ and following
the method of Ref.\ \cite{Mannb}, we obtain
\begin{equation}
p=1+At-Bt\omega -C\omega ^{3}+O\left( t\omega ^{2},\omega ^{4}\right) ,
\label{ptw}
\end{equation}%
where%
\begin{equation}
A=\frac{1}{\rho _{c}},\quad B=\frac{1}{\varepsilon \rho _{c}},
\end{equation}%
and $C$ is%
\begin{equation*}
C=\frac{2(n-1+\alpha ^{2})}{3(1+\alpha ^{2})^{2}\varepsilon ^{3}}-\frac{%
(n-1)(\alpha ^{2}+4n-5)b^{-2\gamma (n-1)/\Upsilon }X^{(n-1)(1-\gamma
)/\Upsilon }}{(\alpha ^{2}+n-2)(\alpha ^{2}+1)^{2}6\beta ^{2}\varepsilon
^{3}q^{2(1-\gamma )/\Upsilon }},
\end{equation*}%
and
\begin{equation*}
C=\frac{(n-1+2\alpha ^{2})}{3\left( 1+\alpha ^{2}\right) ^{2}\varepsilon ^{3}%
}
\end{equation*}%
for large and small $\beta $, respectively. Differentiating Eq.\ (\ref{ptw})
at a fixed $t<0$ with respect to $\omega $, we get
\begin{equation}
dP=-P_{c}\left( Bt+3C\omega ^{2}\right) d\omega .
\end{equation}%
Now, we apply the Maxwell's equal area law\ \cite{Mann1}. Denoting the
volume of small and large black holes with $\omega _{s}$ and $\omega _{l}$,
respectively, we obtain
\begin{eqnarray}
p &=&1+At-Bt\omega _{l}-C\omega _{l}^{3}=1+At-Bt\omega _{s}-C\omega _{s}^{3}
\notag \\
0 &=&\int_{\omega _{l}}^{\omega _{s}}\omega dP.  \label{Equal}
\end{eqnarray}%
Equation (\ref{Equal}) leads to the unique non-trivial solution
\begin{equation}
\omega _{l}=-\omega _{s}=\sqrt{-\frac{Bt}{C}},  \label{oml}
\end{equation}%
which gives the order parameter $\eta =V_{c}\left( \omega _{l}-\omega
_{s}\right) $ as
\begin{equation}
\eta =2V_{c}\omega _{l}=2\sqrt{-\frac{B}{C}}t^{1/2}.
\end{equation}%
Thus, the exponent $\beta ^{\prime }$ which describes the behaviour of the
order parameter $\eta $ near the critical point is $\beta ^{\prime }=1/2.$
To calculate the exponent $\gamma ^{\prime }$, we may determine the behavior
of the isothermal compressibility near the critical point. Differentiating
Eq.\ (\ref{ptw}) with respect to $V$, one obtains
\begin{equation*}
\frac{\partial V}{\partial P}\Big|_{T}=-\frac{V_{c}}{BP_{c}}\frac{1}{t}%
+O(\omega ).
\end{equation*}%
Hence, the isothermal compressibility near the critical point may be written
as
\begin{equation}
\kappa _{T}=-\frac{1}{V}\frac{\partial V}{\partial P}\Big|_{T}\propto -\frac{%
V_{c}}{BP_{c}}\frac{1}{t}\quad \Longrightarrow \quad \gamma ^{\prime }=1.
\end{equation}%
Finally, the shape of the critical isotherm' $t=0$ is given by (\ref{ptw})
\begin{equation}
p-1=-C\omega ^{3}\quad \Longrightarrow \quad \delta ^{\prime }=3.
\end{equation}%
So we have shown that for BI-dilaton black hole in $(n+1)$ dimensions, we
obtain the same critical exponents as in the linear Maxwell case \cite{Mann1}
and in the dilatonic black holes \cite{Kamrani}.

\section{Summary and Conclusions}

In this paper, we investigated the critical behavior of $(n+1)$-dimensional
BI-dilaton black holes in the presence of two Liouville type potentials.
While one of the Liouville type potential guarantees the existence of the
solution, the second one contains a constant $\Lambda $, which has the role
of cosmological constant. We enlarged the phase space by considering the
constant $\Lambda $ and the BI parameter $\beta $ to be treated as
thermodynamic quantities. By calculating the thermodynamic quantities, we
obtained the generalized Smarr relation, which reduces to the Smarr relation
in the absence of dilaton field given in \cite{De,Sherkat1}. After
constructing the Smarr relation, we used the pressure and Hawking
temperature to build the equation of state and plot $P$-$v$ isotherm
diagrams. These figures show the analogy between our system and the van der
Walls fluid, with the same phase transition. We also found that the critical
behavior can occurred only for black holes with spherical horizon ( $k=1$)
provided $\alpha <1$. Then, we obtained the critical pressure, volume and
temperature both for large and small BI parameter $\beta $. We found that
the critical temperature and pressure go to infinity as $\beta$ goes to zero. Indeed, at $\beta =0$ the third
term in Eq. (\ref{Psmall}) vanishes and no critical behavior will occur.  In the absence
of dilaton field ($\alpha =0=\gamma $), our results reduce to those of BI
black holes \cite{Mannb}, while for sufficiently large $\beta $ we recovered
the critical quantities of Einstein-Maxwell dilaton black hole \cite{Kamrani}%
. Moreover, we considered the behavior of the Gibbs free energy and found
that there is a swallowtail behavior for Gibbs free energy as a function of
temperature. This shows that there is a first order small-large black hole
transition in the system. Finally, we calculated the critical exponents and
found that the results are the same as van der Waals system. This, implies
that the inclusion of nonlinear electrodynamics, dilaton field or extra
dimensions do not change the critical exponents. In the present work we only
considered the effects of dilaton field on the critical behavior of the
black holes in BI nonlinear electrodynamics. It is worth to investigate the
effects of dilaton on the critical behaviour of black holes in the presence
of other nonlinear electrodynamic theories.

\acknowledgments{We thank Shiraz University Research Council. This
work has been supported financially by Research Institute for
Astronomy and Astrophysics of Maragha, Iran.}

\end{document}